\documentclass[useAMS,usenatbib]{mn2e}
\usepackage{aas_macros,graphicx,times,multirow,amsmath}
\usepackage[]{color}
\usepackage{subfigure}
\usepackage{latexsym}
\usepackage{txfonts}
\usepackage{hyperref}
\usepackage{amssymb}

\def\be{\begin{equation}}
\def\ee{\end{equation}}
\def\ba{\begin{eqnarray}}
\def\ea{\end{eqnarray}}

\begin{document}

\def\LaTeX{L\kern-.36em\raise.3ex\hbox{a}\kern-.15em
    T\kern-.1667em\lower.7ex\hbox{E}\kern-.125emX}

\def\xmm {\emph{XMM-Newton}}
\def\cxo {\emph{Chandra}}
\def\swift {\emph{Swift}}
\def\sax {\emph{BeppoSAX}}
\def\rxte {\emph{RXTE}}
\def\rst {\emph{ROSAT}}
\def\flux {\mbox{erg cm$^{-2}$ s$^{-1}$}}
\def\lum {\mbox{erg s$^{-1}$}}
\def\v {\upsilon}

\title{Jets and gamma-ray emission from isolated accreting black holes}

\author[M.V. Barkov et al.]{M.V.~Barkov$^{1,2}$, D.V.~Khangulyan$^{3}$ and 
S.B.~Popov$^{4}$\thanks{E-mail: sergepolar@gmail.com}\\
\smallskip\\
$^1$ Max-Planck-Institut f\"ur Kernphysik, Saupfercheckweg 1, 69117 Heidelberg, Germany\\
$^2$ Space Research Institute RAS, 84/32 Profsoyuznaya Street, 117997, Moscow, Russia\\
$^3$ Institute of Space and Astronautical Science/JAXA, 3-1-1 Yoshinodai, Chuo-ku, Sagamihara, 
Kanagawa 252-5210, JAPAN\\
$^4$ Sternberg Astronomical Institute, Lomonosov Moscow State University, 
Universitetsky prospekt 13, 119992, Moscow, Russia  }


\date{}


\maketitle
\begin{abstract}
  Isolated black holes (IBHs)  are not usually considered to be important
  astrophysical sources, since, even in the case of
 a high accretion rate,  an accretion disk rarely can be
 formed due to the small angular momentum of the in-falling
  matter.  Thus, such systems are not expected to feature thermal disk
  emission which makes the dominant contribution to
 the radiative output of binary systems harbouring a BH.  Moreover, due to their
 relatively modest accretion rates, these
 objects are not conventionally treated as feasible jet sources.  However,
  the large number of IBHs in the Galaxy,
 estimated to be $\sim10^8$, implies a very high density of
  $10^{-4}\rm \, pc^{-3}$ and an average distance between IBHs of 
$\sim 10\rm  \, pc$.  Our study shows that the magnetic flux, accumulated on the
  horizon of an IBH  because of accretion of interstellar matter, allows the
  Blandford-Znajeck mechanism to be activated.  Thus,
 electron-positron jets can be launched.  We have performed 2D numerical
  modelling which allowed the jet power to be
 estimated.  Their inferred properties make such jets a feasible
  electron accelerator which, in molecular clouds,
 allows electron energy to be boosted up to $\sim1$~PeV.  For the conditions
  expected in molecular clouds the radiative
 cooling time should be comparable to the escape time. Thus these sources
  can contribute both to the population of
 unidentified point-like sources and to the local cosmic ray
  (CR) electron spectrum.  The impact of the generated electron CRs depends
  on the diffusion rate inside molecular clouds (MCs).  If the diffusion
  regime in  a MC is similar to Galactic diffusion, the produced
  electrons should rapidly escape the cloud and
 contribute to the Galactic CR population at very high energies $>100$~TeV.  
However, due to the modest jet
 luminosity (at the level of $\sim10^{35}\rm \, erg\, s^{-1}$) and low
  filling factor of MC,
 these sources cannot make a significant contribution to
  the spectrum of cosmic ray electrons at lower energies. 
On the other hand, if the diffusion within
  MCs operates at a rate close to the Bohm limit, the CR electrons escaping from the source should be
  confined in the cloud, significantly contributing to the local density of  CRs. The IC emission of these
locally-generated CRs may explain the variety of gamma ray spectra detected from  nearby MCs. 
 
\end{abstract}

\section{Introduction}
\label{intro}

The Galactic population of isolated compact objects -- remnants of
massive star evolution -- is very large: about $10^9$ 
\citep[see, e.g.][and references therein]{st2010}. It should be dominated by
neutron stars, but the number of isolated black holes (IBHs) is also
non-negligible: their number is $\gtrsim 10^8$ \citep{agol2002}.
Taking into account this large number, the local spatial density of IBH
can be high, corresponding to a typical distance to the closest
objects $\sim$tens of parsec.  Such estimates are usually based on
detailed calculations for isolated neutron stars and normalization
based on the relative number of compact objects. The local isolated
neutron star spatial density is estimated to be $\sim 3 \times
10^{-5}$~pc$^{-3}$--$ 3 \times 10^{-4}$~pc$^{-3}$ \citep{petal2005,
  ofek2009, st2010}. The relative number of BHs to 
neutron stars is about 1/3. 
Then, the local density of IBHs can be taken as $\sim
10^{-5}-10^{-4}$~pc$^{-3}$.

Historically, the first ideas about the detectability of IBH were related
to accretion from the interstellar medium (ISM) \citep{sv1971}.
Detailed studies concluded that accreting IBH must emit mostly in
X-rays and IR (see, for example, \citealt{fujita1998}).  Calculations of the
observability of the population of Galactic accreting IBHs were
performed several times (see, for example, \citealt{pp1998} and
references therein).  One of the most detailed studies was presented by
\cite{agol2002}.  Despite several attempts \citep{chisholm2003}, up to
now no firm candidates for accreting IBHs (or for accreting
isolated neutron stars, see \citealt{tr00}) are known.  Optimistically,
future space missions may discover them \citep{exist2009}.

The only possible detections of IBHs reported in recent years are related to microlensing.
About 10 years ago several candidates were reported by different research groups \citep{micro2002, mao2002,
bennett2002}. Recent calculations \citep{st2010} predict that further compact objects will be detected via
microlensing. 


To understand IBHs better, it is very important to seek examples of these not related to microlensing events.
There is a discovery potential for IBHs via modelling of disrupted massive binaries
\citep{pp2002}, or in a close examination of
 unidentified
objects with peculiar properties among  different surveys.  
In this note we develop the idea of
\cite{bk2012} who proposed that
BHs with spherical accretion --- i.e. without formation of an accretion disk
\citep[see also ][]{KB09} --- can be sources of
relativistic jets powered by rotating BH \citep{rw75,lov76} or, e.g., the Blandford-Znajek (BZ) mechanism
\citep{BZ77}.\footnote{In a different
astrophysical context jets from charged IBHs were also studied by
\cite{punsly1998a, punsly1998b,punsly1999, punsly2000,
torres2001}, see also \cite{radio2005}.} 
These jets form conditions for efficient acceleration of non-thermal
particles which can radiate energetic electrons,
with quite low luminosity in X-ray and other energy bands.  Strict
limits on the energy release rate in X-rays
$L_{x}<10^{-3}$ (in units of the Bondi-Hoyle accretion rate with efficiency $0.1c^2$)
was achieved in the work of \cite{mp12}, which is complimentary to the scenario
suggested
by \cite{bk2012}.

In the following two sections we briefly
describe the model. Then, in Sec.\ref{obs} we present our results on the possible observational appearance of IBHs.
Finally,
we present some discussion and our conclusions.

\section{Jets from IBHs}
\label{jets}

Recently, it was shown that the BZ mechanism can be activated in the case of 
direct wind accretion on to a rotating BH in a close binary system \citep{bk2012}.  Thus, a relatively
powerful jet, with kinetic luminosity of 
$\sim 10^{35}\rm \, erg\,s^{-1}$, can be launched in systems which do not feature
an accretion disk.  A similar situation
may arise in the case of accretion of the ISM on to an IBH. 
Accretion from the ISM has been studied for many
decades starting from the seminal papers by \cite{bondi1952, bh1944} (see a
review in \citealt{edgar2004}).  The
case of IBHs  was studied in \citep{sv1971,bkr74,bkr76}
and later this approach was developed in \citep{beskin2005,beskin2008}.

In the simplest case, the accretion rate can be estimated as:
\begin{equation}
\dot{ M}= \lambda 4 \pi \frac{(GM)^2 \rho}{\v^3}.
\label{bondi}
\end{equation}
Here M is the mass of an accretion source, 
$\rho$ is the density of the surrounding medium (the ISM density in our case), and 
$\lambda$ is a dimensionless coefficient $\lesssim 1$.  
The velocity $\v$ must include the contribution from different types of
motion. The most important is just a spatial velocity of the accreting object, $\v_\infty$. When this is small it is
necessary to take into account the sound speed in the medium.  Typically, the sound speed in the ISM is $\sim
10$~km~s$^{-1}$. Spatial velocities are larger, since progenitor stars have typical velocity distribution of 
Gaussian shape with $\sigma \sim 30$~km~s$^{-1}$, and the direction of the BH birth kick is not expected to be related
to the direction of the progenitor star's spacial velocity.  Thus,  in this study we assume that the typical velocities
of IBHs are about several tens of km per second (perhaps as high as hundreds km
per second, similar to the values typical for neutron stars, see \citealt{bhkick2012}), and consequently the additional
contribution due to the ISM sound speed can be neglected.

The rate of matter capture can be estimated from Eq.(\ref{bondi}) as 
\begin{equation}
\dot{M}=7 \times 10^{11} \lambda \, M_{1} \v_{1.5}^{-3}
\rho_{-24}~{\mathrm{g~s}}^{-1}, 
\label{mdot}
\end{equation}
here $M_{1}=M/10\,M_{\odot}$, $\v_{1.5}=\v/10^{1.5}$~km~s$^{-1}$~
$\approx \v/31$~km~s$^{-1}$,
$\rho_{-24}=\rho/10^{-24}$~g~cm$^{-3}$ are the mass of an IBH, the
peculiar velocity of the IBH and the density of the ISM,
respectively \footnote{In this paper we use the prescription $A_x=A/10^x$.}. Since IBHs are not expected to be
powerful sources, we mostly focus on the most nearby locations \citep{boch90,slcw99,maiz01}. This allows the ISM density
to be estimated. Namely, currently the solar system is located in the {\it Local interstellar cloud}, which is expected
to have particle density of $\sim0.1$~proton~cm$^{-3}$, or  equivalently $\rho_{\rm lic}\sim 2\times10^{-25}\rm \,
g\,cm^{-3}$. The surrounding region, the so-called the {\it Local Bubble}, is characterized by a much smaller density,
$\rho_{\rm lb}\sim 10^{-26}\rm \, g\,cm^{-3}$. Finally, at distances $d_{\rm mc}\geq150\rm\, pc$ from the solar system
there is a number of {\it molecular clouds} (MC) 
which may have rather high density $\rho_{\rm mc}\sim 10^{-21}\rm \, g\,cm^{-3}$.
Therefore, unless the BH proper velocity differs significantly from $\v_{1.5}\sim 1$, the expected accretion rates for
the LIC, LB and MC are $\dot{M}_{\rm lic}\simeq 10^{11}\rm \,g\,s^{-1}$, $\dot{M}_{\rm lb}\simeq 10^{10}\rm
\,g\,s^{-1}$, and $\dot{M}_{\rm mc}\simeq 10^{15}\rm \,g\,s^{-1}$, respectively. 

In the cases that the accretion regime differs from  Bondi-Hoyle accretion, the rate can be even smaller. In
particular,  for low velocity IBHs, $\v\lesssim 50$~km~s$^{-1}$, a temporary disk can be formed due to interstellar
turbulence  \citep{fujita1998}. The accretion rate in this case can be 
slightly smaller than the Bondi value \citep{bb99} due to the
formation of a wind outflow from the accretion disk. However, such   disks are expected to be transient on a time-scale 
$t_{\rm bdt}\sim R_{ \rm B}/\v_\infty$, where $R_{ \rm B}$ is the so-called Bondi radius:
\begin{equation}
R_{ \rm B}=2GM/\v_\infty^2= 2.5 \times 10^{14} M_1 \v_{1.5}^{-2}\, {\rm cm}
\label{rb}
\end{equation}
and
\begin{equation}
t_{\rm bdt}\sim R_{ \rm B}/\v_\infty \approx 2GM/\v_\infty^3 \approx  10^{8} M_1
\v_{1.5}^{-3}\, {\rm s}.
\label{tbdt}
\end{equation}
Thus, for realistic cases, this time-scale is very short, $\sim 1$~yr. Therefore in what follows we adopt the values 
obtained for the Bondi accretion regime. 

In the model by \cite{bk2012} a jet is driven by the BZ mechanism. The total power in the jet can be estimated by
a simple numerical relationship 
 \begin{equation}
L_{ \rm BZ} \approx C\dot{M}c^2 = 10^{31} \dot{M}_{11} C_{-1} \mbox{ erg s}^{-1},
\label{lbzn}
\end{equation}
here $C_{-1}=C/0.1$ is a non-dimensional parameter $C_{-1}\leq 1$. The maximum value of  $C_{-1}\simeq 1$ can be
achieved 
if high enough magnetic flux can be accumulated at the BH horizon.  The required magnetic field flux can be obtained
through the following relation \citep{BK08b}:
\begin{equation}
L_{ \rm BZ}=1.4 \times 10^{29} f(a) \Psi_{17}^2 M_1^{-2} {\rm erg}\,{\rm s}^{-1},
\label{lbzp}
\end{equation}
here $\Psi_{17}=\Psi/10^{17}$ is magnetic flux in units $10^{17}$~G~cm$^{2}$.
The function$f(a)=a^2\left(1+\sqrt{1-a^2}\right)^{-2}$ is a dimensionless
function accounting for the BH rotation\footnote{Let us note that 
a BH can be significantly 
slowed down if it had accreted
about 0.1 of its mass. 
In the case of IBHs using Eq.(\ref{mdot}) it will last up to $t_{\rm sd}\sim 0.1 M_{\rm
BH}/\dot{M}\sim 10^{14} M_{1} \dot{M}_{12}$~years, so we can neglect the IBH braking.}.  Combining 
Eqs.(\ref{lbzn}) and (\ref{lbzp}) and adopting $f(a)\approx 0.1$  (or $a\sim 0.5$), one can  derive the magnetic flux 
value on BH horizon, required for the optimal operation of the BZ mechanism:
\begin{equation}
\Psi_{\rm BZ} \approx 3\times 10^{18} \dot{M}_{11}^{1/2}  M_1 \mbox{ G cm}^2.
\label{psibz}
\end{equation}
%
%
The value of $\Psi$ which an IBH can accumulate by Bondi accretion is  
$\Psi_{\rm B} = 2 R_{ \rm B}^2 B_{\rm ISM}$ or 
\begin{equation}
\Psi_{\rm B} \approx 3 \times 10^{22} R_{\rm B,14}^2 B_{\rm ISM,-6} \mbox{G cm}^2.   
\label{psiism}
\end{equation}
Therefore, in the case of accretion of the ISM matter on to an IBH the BZ mechanism can achieve its optimal 
performance provided by Eq.~(\ref{lbzn}). However, we note that there are some uncertainties related to role of
the reconnection and magnetic flux escape from the horizon of an IBH. 

Thus, for the expected accretion rates, 
the IBH might be characterized by a jet with a luminosity of 
$L_{\rm j,lic}\sim10^{31}\rm \, erg\,s^{-1}$, $L_{\rm j,lb}\sim10^{30}\rm \, erg\,s^{-1}$ 
and $L_{\rm
j,mc}\sim10^{35}\rm \, erg\,s^{-1}$ for the cases of BHs located in LIC, LB and
MC, respectively. Given the dependence of
the jet luminosity on different factors, e.g., the accretion rate, on the magnetic field
at the horizon, spin and
velocity of the IBH, these luminosities can be, in fact, considered as upper limits.

\section{Spatial density of IBHs}
\label{sdibh}
Another important issue is related to the actual number of such IBHs in the
ISM with different properties.  For
the above estimates there should be 
$N_{\rm IBH}=4\times(10^{-1}-10^{-2})R_{\rm 1}^3$ located within
$R=10R_{\rm 1}$~pc from the solar system.  For the expected 
size of the LIC of $\sim10\rm\,pc$, this yields a probability of $4\%-40\%$
for a BH to be located there.  With increasing
$R$ the number of BHs  rises rapidly achieving $10^{3-4}$ inside $\sim
1$~kpc.  On the other hand, IBHs in high density regions are rare: they are
at least two orders of magnitude less frequent than IBHs in the normal and
low-density ISM.  A typical MC with $M\sim10^5\, M_\odot$ has a volume $\sim 10^4$~pc$^3$. 
I.e., for the IBH spatial density close to the Galactic plane 
$\sim 10^{-4}$~pc$^{-3}$, we expect one IBH per cloud. 
On larger scales, the number of IBHs located in MCs might be very significant.  Given, however, the presence of pulsars
on this scale, the impact of IBHs is negligibly small.  For example, the Vela pulsar is located at a distance of 290~pc,
and is expected to be a powerful source of cosmic ray (CR) leptons with power comparable to its spin-down
luminosity, $L_{\rm sd}=7\times10^{36}\rm erg\,s^{-1}$ \citep[see e.g.][and references therein]{hfp11}.
The closest IBH in a cloud is expected at a distance $\sim 170$~--~200~pc (see the list of near-by MCs, for example in
\citealt{dame1987}).  Up to $\sim 300$~pc  only $\sim5$ IBHs are expected to be located in MCs.

\section{Observational appearance of IBHs jets}
\label{obs}
\subsection{Particle acceleration}
Non-thermal processes in BH jets can be linked to shocks  
at the head of the jet. A system consisting of two
shocks is expected to be formed: 
a forward shock propagating through the ISM and a relativistic reverse
shock in the jet matter. These shocks are characterized by very different velocities.
Properties of the non-thermal particles accelerated by these shocks may differ significantly therefore. 
The jet power and medium density determines
the key properties of the forward  shock.  Namely, the jet ram pressure should be balanced by the inertia of the ISM.
This yields the following relation $L_{\rm j}=p_{\rm ram}dV/dt = p_{\rm ram} \pi R_{\rm S}^2 c$, where $p_{\rm ram}\sim
\rho \v^2$  and $R_{\rm S}$ are the jet ram pressure and the radius of the termination shock, respectively. The
latter can be
estimated as
\be
R_{\rm S}\simeq3\times10^{15} L_{\rm j,31}^{1/2}\rho_{-24}^{-1/2}\v_{1.5}^{-1} \rm cm\,.
\label{eq:shock}
\ee
An upper limit on the forward shock magnetic field strength can be derived from pressure balance  $B_{\rm S}^2/8\pi
\lesssim 
p_{\rm ram}$:
\be
B_{\rm S}\simeq15\rho_{-24}^{1/2}\v_{1.5}\mu \mathrm{G}\,.
\label{eq:forward_B}
\ee

The reverse shock in the jet has a similar size and, in fact,  its magnetic field strength should be comparable.
Indeed,  since jets launched by the BZ mechanism are magnetically dominated, the magnetic field strength can be
estimated as 
\be
B_{\rm j}=\sqrt{4L_{\rm j}\over c}\frac1{R_{\rm S}}\,.
\ee
This allows one to obtain the following relation:
\be
B_{\rm j}R_{\rm S}=10^{11} M_1^{1/2}\v_{1.5}^{-3/2}\rho_{-24}^{1/2}C_{-1}^{1/2}\rm \, G\,cm\,.
\ee
For the shock size $R_{\rm S}$ determined by Eq.(\ref{eq:shock}), this yields a magnetic field of
\be
B_{\rm j}=10 \rho_{-24}^{1/2}\v_{1.5} \mu \mathrm{G}\,,
\label{eq:b_j}
\ee
which is quite close to the strength expected at the forward shock Eq.(\ref{eq:forward_B}). The main difference between 
these two shocks is related to the efficiency of the acceleration process.
 
The acceleration time of non-thermal particles can be estimated as 
\be
t_{\rm acc}={\eta R_{\rm g}\over c}\,
\ee
here $R_{\rm g}$ is electron giro-radius, and $\eta>1$ is a dimensionless parameter, which in the case of diffusive
shock acceleration has a value of  $\eta_{\rm diff}={2\pi \left({c /\v_{\rm sh}}\right)^2}$. In the case of the
forward termination shock, it is natural to assume that $\v_{\rm sh}\simeq \v$. In the case of the reverse shock, which
is expected  be characterized by relativistic velocities,  the value of $\eta$-parameter is not defined in
the framework of
theoretical models. However, from the interpretation of the non-thermal emission detected from the Crab Nebula, one can
infer  that in the relativistic case $\eta < 100$, thus in what follows we will assume $\eta_{\rm rel}=10$.

In the case of diffusive shock acceleration the maximum energy of the non-thermal particles can be limited either
 by loss rate or by the confinement requirement. The latter is refereed to as the {\it Hillas criterion} \citep{hill84},
which for the Bohm diffusion regime (i.e., the diffusion coefficient is $D\sim R_{\rm g} c$) can be formulated as a
limitation of the size of the acceleration site  $R_{\rm acc}$:
\be
R_{\rm acc}> R_{\rm g} \sqrt{\eta}\,.
\ee
In the asymptotic case $\eta=1$ so this condition results in an obvious requirement of $R_{\rm acc} > R_{\rm g}$. Since 
in the case of the jet termination $R_{\rm acc}\sim R_{\rm S}$, the electrons can be accelerated up to 
\be
E_{\rm max, f} <5 C_{-1}^{1/2}\rho_{-24}^{1/2}M_1^{1/2}\v_{1.5}^{-1} \rm GeV \,.
\label{eq:emax}
\ee
This value gives $\sim2$~GeV for the Local cloud, and $\sim0.5$~GeV for a BH located in the
Local Bubble. In the case of a BH in a MC, the maximum energy achieved at the jet termination shock is $\sim
200$~GeV. Therefore, the jet forward termination shock is not a plausible acceleration site for any reasonable
conditions.

Regarding the reverse shock, the Hillas Criterion gives the following estimate: 
\be 
E_{\rm max, r}< 30 C_{-1}^{1/2}\rho_{-24}^{1/2}M_1^{1/2}\v_{1.5}^{-3/2}\eta_{1}^{-1/2}\rm \, TeV\,.  
\ee 
This gives $\sim10$~TeV for the Local cloud, and $\sim 1$~TeV for BH located in the Local Bubble. In the case of
an IBH in a MC, the maximum energy of particles accelerated at the jet termination shock is $\sim 1$~PeV.

\subsection{Source emission}

{ The acceleration process is limited by the Hillas Criterion, therefore
the particles are not expected to loose all their energy in the
source.  Nevertheless, we compare the radiation cooling time to the
particle escape time.  The latter depends on the regime of 
diffusion inside the source. Obviously, this diffusion rate is
different from the Galactic diffusion coefficient used to derive
Eq.~(\ref{sdif}) and it cannot be determined from first
principles. However, the X-ray properties of supernovae remnants
apparently indicate that inside these shock related non-thermal
sources, the diffusion regime is similar to the Bohm limit
\citep{fabook04}. We therefore assume this diffusion regime to be
valid inside the source. Thus, the source escape time can be estimated
as }
\be
t_{\rm bd}=5\times10^{7}E_{\rm TeV}^{-1}\v_{1.5}^{-4}\rho_{-24}^{1/2}\,\rm s\,.
\label{eq:diff_source}
\ee
Moreover, we note that the assumption of Bohm diffusion inside the source is in fact
closely related to the acceleration rate assumed above.

The radiative cooling time is determined by the density of the
corresponding target. In the case of the synchrotron and inverse
Compton (IC) radiation mechanisms on the magnetic field
$\varepsilon_{B}=B^2/8\pi$ and photon $\varepsilon_{\rm ph}$ energy
densities: 
\be
t_{\rm ic}=3\times10^5 (\varepsilon_{\rm ph}/1 {\rm
  eV\,cm^{-3}})^{-1} E_{\rm TeV}^{-1}\rm\, yr\,
\label{tic}
\ee
and the synchrotron cooling 
\be
t_{\rm syn}=4\times10^5 (B/5{\rm \mu G})^{-2} E_{\rm TeV}^{-1}\rm\, yr\,.
\label{tsyn}
\ee
{
Equations~\ref{eq:forward_B} and \ref{eq:b_j} allow the synchrotron
cooling time to be obtained, and the IC cooling time can be estimated
based on the size of the termination shock. Jet kinetic energy is
largely transformed to the thermal energy at the termination
shock. The photon energy density inside the source can be estimated as
$\varepsilon_{\rm ph} \approx L_{\rm j}/2\pi R^2_{\rm S} c\sim
3\rho_{-24} \v_{1.5}^2 \mbox{ eV cm}^{-3}$, which is
comparable to the magnetic field energy density:
$ \varepsilon_{\rm ph} \sim \varepsilon_{B}  \sim 5\rho_{-24}
\v_{1.5}^2 \mbox{ eV cm}^{-3}$. 
Thus,  assuming that IC scattering operates in the Thomson
regime we can estimate the radiative cooling time as: }
\be 
t_{\rm
  rad}=1.5\times10^{12}\rho_{-24}^{-1}\v_{1.5}^{-2}E_{\rm
  TeV}^{-1}\rm\, s\,.
\label{eq:cooling}
\ee
The lower energy part of the spectrum, $E< 1\rm \,GeV$, is dominated by bremsstrahlung losses 
\be 
t_{\rm bs}=6\times10^{7} \rho_{-24}^{-1}\rm \, yr\,. 
\label{tbs}
\ee

The non-thermal particles emit in the source a fraction of their energy, which depends on the
ratio of the diffusion and radiative cooling times:
\be
L_{\rm ph}\sim{t_{\rm bd}/t_{\rm rad} \over 1+t_{\rm bd}/t_{\rm rad}}\chi L_{\rm j}\,,
\label{eq:nonthermal}
\ee
where $\chi$ is the fraction of the jet kinetic luminosity which is transferred to the population of the non-thermal
particles. The estimate Eq.~(\ref{eq:nonthermal}) depends strongly on the density of the surrounding medium:
\be
{t_{\rm bd}\over t_{\rm rad}}=3\times10^{-5}\rho_{-24}^{3/2}\v_{1.5}^{-2}\,.
\label{eq:nonthermal2}
\ee
It can be seen that, for the case of the density of the medium expected in MCs, i.e., $\rho\sim10^{-21}\rm \,
g\,cm^{-3}$, this estimate approaches unity. 
Thus, while in the cases of {\it Local Bubble}  and {\it Local interstellar
  cloud}, jets emerging from IBHs are not expected to be plausible point sources, in the case of an IBH accreting in
a MC,  non-thermal particles can emit a significant part of their energy in the close vicinity of the source. One should
expect similar fluxes emitted through the synchrotron and IC channels, since the energy densities of the target fields
are similar. However, these components are located in different energy bands. The synchrotron emission should be emitted
at an energy
\be
\epsilon_{\rm syn}=10 \v_{1.5}\rho_{-21}^{1/2} E_{\rm TeV}^2 \,\rm eV\,,
\label{eq:synch}
\ee
where $E_{\rm TeV}$ and $\rho_{-21}$ are the energy of the emitting particle in TeV units and density of the MC in
$10^{-21} \rm\, g\,cm^{-3}$ units. For the 
assumed condition at the reverse shock electron energy can
significantly exceed 1~TeV (see Eq.~\ref{eq:emax}), thus the synchrotron emission can cover quite a broad range from
radio to soft gamma-rays.

The main target for the IC scattering is provided by the shocked MC material, which is expected to be heated to $\sim50\rm
\, K$.  In the Thomson limit the energy of the up-scattered photon can be estimated as:
\be
\epsilon_{\rm ic}=20 E_{\rm TeV}^2 \,\rm GeV\,.
\label{eq:synch}
\ee
{ The transition to the Klein-Nishina regime occurs for electrons of
energy $\gtrsim20$~TeV, i.e. in the band which is slightly above the
range of  optimal sensitivity of modern Cherenkov
detectors. Thus, in what follows we do not discuss this energy range.
It is also worthy to note that the spectral shape is expected to be
flat with photon index $\sim2$, without any significant
breaks. Indeed, since the dominant cooling processes -- synchrotron
and IC -- and escape time have the same energy dependence, the spectral
breaks can be induced either by the transition to 
bremsstrahlung cooling dominance, or by the age of the source. The later is
not relevant for sources with age $>10^4$~yr, and the bremsstrahlung
break should appear in the synchrotron spectrum at energy of 
  $<1$~GHz, i.e. in the range, which can hardly be probed with observations.

However, if the diffusion deviates significantly from
the Bohm regime, this may lead to modification of the spectral shape,
and a break in the electron spectrum should appear at energy
corresponding to $t_{\rm esc}=t_{\rm rad}$. Also a weak Klein-Nishina
hardening \citep[see e.g.][]{ka05,msc05} is expected to appear in the
X-ray energy band $\epsilon_{\rm syn}\sim10\rm\, keV$.
}


\subsection{Particle cooling and diffusion in ISM}

Since even in the case of IBHs located in MCs a significant fraction of the non-thermal power is
able to escape from the
source, below we study whether these sources 
can produce a detectable contribution to the diffuse background. 
However, the total
power of these sources per kpc$^3$ is $\sim10^{36}\rm erg\,s^{-1}\,kpc^{-3}$; this is approximately 2 orders
of magnitude below the rate required to power the CR electron spectrum \citep[see e.g.][]{fabook04}. Therefore, these
sources can give an important contribution only in the high energy part of the spectrum, where more distant sources are
heavily suppressed due to particle energy losses.

Following previous studies  \citep{pms08,smr00,daf97} we can say that the 
Galactic background radiation field in the solar vicinity
 is determined by 3 main components: 1) cosmic microwave background (CMB) ($u_{cmb}=0.25 \rm\, eV\,cm^{-3}$); 2) far
infra-red (FIR) radiation of a dust ($u_{\rm fir}=0.3 \rm\, eV\,cm^{-3}$) 
and 3) visible light radiation from stars ($u_\mathrm{op}=0.4 \rm\,
eV\, cm^{-3}$). We should point out that for optical radiation, lepton with $E>1 \rm TeV$
interact in the Klein-Nishina regime, this  significantly suppressing the interaction cross-section.   
The energy density of the magnetic field near the Sun is $u_B=B^2/8\pi=0.6 \rm\, eV\, cm^{-3}$ \citep[see,
e.g.,][]{boch90}. The total energy
density can be determined as $u_{\rm r+B}= u_{\rm fir}+u_{\rm cmb}+u_{\rm B}
+u_{\rm op} \approx 1.2-1.5 \rm\, eV\, cm^{-3}$.

The cooling time scales (Eqs.~(\ref{tic}-\ref{tbs})) should be compared to the diffusion time 
\be 
t_{\rm dif}=10^5 R_{2.5}^2 E_{\rm  TeV}^{-\delta}\rm\, yr\,.  
\label{sdif}
\ee 
Here $R_{2.5}=R/310$~pc and $\delta\approx 0.5$. This diffusion time is normalized to
the distance to the closest young energetic pulsar (the Vela pulsar) which is expected to be a powerful source of CR
electrons \citep{hfp11}. A comparison between $t_{\rm ic}$ and $t_{\rm dif}$ allows one to determine the energy range
for which the nearby sources can give a more important contribution than pulsars. Namely, a cooling cut-off is expected
at
energy 
\be 
E_{\rm cut}\approx 5 R_{2.5}^{-2/(1-\delta)} \rm \,TeV.  
\ee 
In the case of IBHs (R=170~pc, see Eq.~\ref{sdibh}) $R_{2.5}=0.56$ and we get an exponential cut-off near 50 TeV. Taking
into account that the relation between pulsar and IBH power is $\xi \approx 100$, we can estimate the energy range where
IBHs should dominate the spectra of CR electrons and positrons as $E_{\rm br}\approx E_{\rm cut}
\log(\xi)^{1/(1-\delta)}\approx 100$~TeV.
On the other hand, these energies are close to the limiting value for IBHs. 
 If however multi-TeV CR electrons are detected (e.g., with Cherenkov
detectors, see \citealt{hesses08}), the IBH scenario could provide a feasible explanation. { Future experiments
like CTA can shed light on this problem. }

Regarding a possible contribution to CR positrons, IBHs can supply particles  in  the high energy band (GeV
and TeV).  However, their contribution is expected to be significantly exceeded by the contribution from the nearby
pulsars, e.g. from the Vela pulsar.

\section{Discussion and conclusions}
\label{disc}

In this paper we examine IBHs as possible jet sources. We find that the non-thermal efficiency of these sources depends
strongly on the density of the surrounding medium. IBH jets launched in high density environments are characterized
by a larger power and a stronger magnetic field. The latter feature appears to be very important since (1) it allows
non-thermal particles to be accelerated up to VHE; and (2) the ratio of the escape time to radiative cooling time
decreases with increase of the magnetic field.  The key source parameters (jet luminosity, maximum energy of
non-thermal particles and non-thermal luminosity)  scale with density as: $L_{\rm j}\propto \rho$, $E_{\rm
max}\propto \rho^{1/2}$ and $L_{\rm ph}\propto \rho^{5/2}$.  The strong dependence of the non-thermal power, $L_{\rm 
ph}$, on the medium density is related to the change of the ratio of radiative to escape losses and, in the case of
high enough density, converges to the {\it fast cooling regime}, where $L_{\rm ph}\propto L_{\rm j}\propto\rho$.

More specifically, if the density of the surrounding medium is high, $\rho\gtrsim10^{-21}\rm \, g\, cm^{-3}$, the
radiation cooling time is shorter than the escape time (if the Bohm diffusion
regime is applies within the source). Thus, the jet reverse shock can turn into an efficient emitter of non-thermal
radiation. On the other hand, if the density is lower, $\rho\lesssim10^{-21}\rm \, g\, cm^{-3}$, the IBH should act as a
source of electron CR, interestingly, for conventional values of the MC density ($\rho\sim10^{-21}\rm \, g\,
cm^{-3}$), the non-thermal power of these sources should be shared in comparable fractions between the radiative cooling
inside the source and escape.  Finally, for the IBHs accreting in low density environments $\rho\ll10^{-21}\rm \, g\,
cm^{-3}$, the accretion rate and maximum available energy of non-thermal particles appears to be too low to make any
detectable contribution to the CR spectrum.  Moreover, as fast particles escape these sources, 
then they should not be considered as feasible point-like emitters.


As shown above, in the case of accretion in high density medium, the jets launched by
IBHs, may emit a significant
fraction of the accreted energy through synchrotron (from radio to soft gamma-rays) and IC channels (up to very high
energies) within  $10^{16}\rm\, cm$ from the jet termination point. The luminosities radiated via these two mechanisms
should be similar, since the energy densities of the target fields are expected to be comparable.  The dominant target
photon field for the IC is expected to be provided by shocked MC matter, which should be heated to $T\sim50\rm \,
K$. Because of this relatively low temperature of the target field, the Klein-Nishina effect should not affect the shape
of the IC spectrum in the range relevant for the modern gamma-ray instruments. However, the Klein-Nishina effect may
lead to a weak hardening of the synchrotron X-ray component \citep[see, e.g.,][]{ka05,msc05}. Thus, the spectra are
expected to be featureless, with a photon index of $\sim2$ over a broad energy range. Any stronger spectral
features in the spectrum would be attributed to a deviation of the escape rate from the Bohm diffusion regime. It is
also important to note that one can expect a certain variability of the non-thermal flux caused by the
inhomogeneous structure of MCs \citep[see, e.g.,][]{kp79}. For a typical velocity of $30\rm\,km\,s^{-1}$, these
inhomogeneities should lead to variability on a year time scale.  

For the expected non-thermal luminosity of $<10^{35}\rm\, erg\,s^{-1}$, such sources should be detectable with modern X-
and gamma-ray instruments up to a distance of $\sim1\rm\,kpc$. Within this distance $\sim10$ IBHs should
be located in
MCs. This may provide a feasible interpretation for some unidentified gamma-ray sources.  




Regarding the impact of the non-thermal particles escaped 
from the shock region generated by IBHs in MCs, there are
certain uncertainties related to the diffusion of these particles. Namely, the propagation of such particles may proceed
very differently within and outside the MC.  If the diffusion coefficient in the MC is comparable to the Galactic one,
then the high energy particles should rapidly escape from the MC (the MC crossing time is $\sim 300\rm \,yr$). However,
several uncertainties do not allow us to claim IBHs to be the best candidates for production of very high cosmic ray
electrons. The distribution of spins of IBHs is unknown and if $a<0.3$ jets will be weak or simply not
launched at all \citep{KB09,bk10}. Nevertheless, we have checked the contribution from the nearby pulsars up to the
distance 300~pc  using ATNF (The Australia telescope national facility) pulsar catalogue \citep{mht05}. We can conclude
that the contribution in the CR electron spectrum from close pulsars is about a factor 10
smaller than expected contribution from IBHs.  Although we can miss some pulsars if their beamed radiation does not
cross the Earth, such pulsars can contribute to the population of electrons near the Sun. So far it is unlikely that
IBHs are the main source of electrons in CR, except hypothetical particles with $E>100$~TeV.

However, the physical scenario should be quite different if in MCs the diffusion operates slower. For example, for
the Bohm diffusion regime, the particles confinement in MC should be very long, $\sim10^7\rm\,yr$. Thus, the non-thermal
particles injected by a IBH should produce a diffusive emission component associated with the host MC. The luminosity of
this component should be comparable to the non-thermal power of the source, i.e. $<10^{35}\rm\, erg\,s^{-1}$, so it
might give a contribution comparable to the fluxes produced by CR interacting with the MC material
\citep{aharon01,caf10}. Importantly, the spectra produce by CR protons interacting with the MC should be quite different from
the IC spectra generated by the local population of CR electrons. Thus, this scenario can contribute to our
understanding of the variety of the GeV spectra obtained from different MCs with {\it Fermi}/LAT (Yang and Aharonian, in preparation).

\section*{Acknowledgments}
The authors thank Felix Aharonian, Valenti Bosch-Ramon and Maxim Pshirkov for useful discussions,
Roland Crocker for help in preparation of the paper and the anonymous referee for useful  suggestions.
The work of S.P. was supported by
RFBR grants (10-02-00599, 12-02-00186) and by the Federal programme for
scientific and
educational staff (02.740.11.0575).


\end{document}